\input epsf 
\documentstyle[preprint,eqsecnum,aps]{revtex}

\begin{document}

\count255=\time\divide\count255 by 60 \xdef\hourmin{\number\count255}
  \multiply\count255 by-60\advance\count255 by\time
 \xdef\hourmin{\hourmin:\ifnum\count255<10 0\fi\the\count255}

%\draft
\preprint{WM-01-106}

\title{Low-scale Quantum Gravity and Double Nucleon Decay}

\author{Carl E. Carlson\footnote{carlson@physics.wm.edu} and 
Christopher D. Carone\footnote{carone@physics.wm.edu}}

\vskip 0.1in

\address{Nuclear and Particle Theory Group, Department of
Physics, College of William and Mary, Williamsburg, VA 23187-8795}

\vskip .1in
\date{March, 2001}
\vskip .1in

\maketitle
\tightenlines
\thispagestyle{empty}

\begin{abstract}
In models with a low quantum gravity scale, one might expect 
sizable effects from nonrenormalizable interactions that violate the
global symmetries of the standard model.  While some mechanism must
be invoked in such theories to suppress higher-dimension operators that 
contribute to proton decay, operators that change baryon number by two 
units are less dangerous and may be present at phenomenologically 
interesting levels.  Here we focus on $\Delta B=2$ operators that also 
change strangeness.   We demonstrate how to compute explicitly a typical 
nucleon-nucleon decay amplitude, assuming a nonvanishing six-quark cluster 
probability and MIT bag model wave functions.  We then use our results to 
estimate the rate for other possible modes.  We find that such 
baryon-number-violating decays may be experimentally accessible if 
the operators in question are present and the Planck scale is less than 
$\sim$ 400 TeV.
\end{abstract}


\newpage
\setcounter{page}{1}

\section{Introduction}\label{sec:intro}

One of the exciting implications of extra spacetime dimensions 
with large radii of compactification is the possibility that the scale
of quantum gravity may be brought down to TeV energies~\cite{lps}.  
While much effort has focused on understanding the experimental signals of 
the extra dimensions themselves (for example, through the effects of 
Kaluza-Klein excitations in the effective four-dimensional theory), 
much less has been said on the physics that originates at the cut 
off~\cite{fcncs}.  This is understandable for two reasons.  First, any 
detailed knowledge of Planck-suppressed operators in a theory with a low 
quantum gravity scale requires a complete theory of quantum gravity, which 
is not yet at hand.  Alternatively, a general effective field theory 
approach, in which one includes all operators consistent with the gauge 
symmetries of the standard model and suppressed by powers of the cut off, 
leads to baryon- and lepton-number-violating effects far in excess of 
the experimental bounds~\cite{gko,aranda}.  The most common approach to this 
dilemma is to assume that some mechanism forbids the undesirable 
operators, and then to ignore the issue altogether.  Here we will explore 
the possibility that the mechanism responsible for maintaining proton 
stability does not forbid the complete set of operators that violate the 
global symmetries of the standard model. Assuming $\Delta B= 1$ interactions 
are absent, higher-dimension operators that violate baryon number $B$ by two 
units are far less problematic, and may be present in low Planck scale 
scenarios at a phenomenologically interesting level.  Moreover, such 
operators are not suppressed by separating quarks and leptons in an extra 
dimension, as has been suggested as a remedy to the proton decay 
problem~\cite{ns}. In this letter we study the two-body double 
nucleon decays that are induced by $\Delta B=2$ operators and determine the 
sensitivity of existing experiments to the scale of the new physics.

The idea that the most dangerous baryon-number-violating operators
may be absent, while others are present is not at all a radical one.
Consider any model in which  baryon-number is promoted to a
gauge symmetry and then spontaneously broken: higher-dimension operators 
that violate baryon number by $\Delta B$ units are induced in the low-energy 
theory, but the smallest possible value of $\Delta B$ is controlled by
the charge of the Higgs field that is responsible for the spontaneous 
symmetry breaking~\cite{aranda}.  From a low-energy perspective, the 
original continuous gauge symmetry is irrelevant, and one concludes that 
a discrete remnant is sufficient to eliminate the unwanted interactions.  
Such ``discrete gauge symmetries'' are known to be preserved by quantum 
gravitational effects~\cite{kw}, are well defined as fundamental 
symmetries~\cite{bnks}, and arise in string theory~\cite{bd}.  It is not 
hard to imagine scenarios in which operators that contribute to nucleon 
decay are forbidden by some residual discrete symmetry below the string 
scale, while operators of higher-dimension that violate baryon number 
remain in the low-energy theory~\cite{fn1}.

Past interest in effective $\Delta B= 2$ interactions has appeared in
the context of grand unified theories~\cite{guts}, R-parity-violating 
supersymmetry~\cite{sg}, and Planck-scale physics~\cite{benakli}. The 
relevant dimension-nine operators have been cataloged in the 
literature~\cite{wolfenstein,shrock}, but have not been 
studied in their entirety.  This is due in part to the difficulty in 
evaluating hadronic matrix elements, and the relatively large number of 
operators involved. The fact that each operator has an undetermined 
coefficient of O(1) leads to an unavoidable theoretical uncertainty, and 
makes the value added in undertaking a complete analysis somewhat small.  
We will proceed by selecting a typical operator and decay process that is
convenient for an {\em explicit} matrix element evaluation (in fact, one 
that has never appeared in the literature); we then use this result to 
estimate the size of other accessible modes.  

Our canonical decay calculation will be for the process $D \rightarrow K^* K$,
where $D$ represents a deuteron. We choose this mode for a number of 
reasons: (i) The underlying operator changes strangeness and is unconstrained 
by neutron-antineutron oscillation bounds. (We will have more to say about 
the relationship between the operator we consider and others that do not 
change strangeness in the final section.) (ii) the operator we consider
contributes to the matrix element of this process via precisely one
Feynman diagram, and (iii) the spin-flavor-color-spatial wave function of 
the initial state is easily cross-checked with studies of the deuteron 
structure that appear in the literature~\cite{ms77}.  We compute the 
effective lifetime for this decay, and then extrapolate to other NN modes of 
interest.  We will show that such two-body decays may be accessible if 
the Planck scale is less than $\sim 400$~TeV, and we discuss the dependence 
of our result on the flavor structure of the theory.

How then can a significant number of quarks within a deuteron be annihilated 
by a contact interaction, if the quarks are spatially separated within 
the neutron and proton?

\section{Deuteronomy}
A favorable answer to this question is that the deuteron is not entirely
made of a proton and neutron, but also includes a significant admixture
of a six-quark (6q) cluster state~\cite{ms77}.  In the case at hand, the
6q cluster is a  six quark state that is totally antisymmetric in color,
spin, and flavor, with isospin~0 and spin~1.  Such a state is not
factorable into a simple product of two color-singlet, three-quark states.

When the constituents of the nucleus are far apart,  a description in
terms of neutrons and protons is accurate.  The question is what happens
to the material inside the nucleus when the pieces come close to each
other.  The quark cluster viewpoint is that if two nucleons come sufficiently
close together, the quarks within them reorganize into a new state where
each quark is in the lowest energy spatial state,  and the
color-spin-flavor part of the quark wave function is uniquely fixed~\cite{ms77}
by the requirement that it be totally antisymmetric, colorless, and of the
desired spin and isospin.

Quark clustering gained impetus~\cite{pv81} in explaining $^3$He data in
kinematic regions inaccessible to scattering off a stationary nucleon and where
contributions due to Fermi motions calculated in standard models were
insufficient.  Further, differences between quark distributions in 6q clusters and
in nucleons provide one straightforward explanation~\cite{ch83} of the nuclear
EMC effect.

The deuteron state may be expressed as a linear combination

\begin{equation}
\big|  D  \big\rangle   = 
(1-f)^{1/2} \big|  D, NN  \big\rangle + f^{1/2} \big| D, 6q \big\rangle \ , 
\end{equation}

\noindent where $f$ is the 6q cluster probability.  The crucial point is
that there is no reason to expect the overlap of the quark spatial wave
functions to be negligibly small in a 6q cluster.  Providing $f$ is also
non-negligible, we avoid the possibility of obtaining an uninteresting
result due to wave function suppression.  

Values for $f$  have been estimated from 6q cluster models of the
nuclear EMC effect~\cite{ch83}; from calculations of the probability for 
nucleons to overlap  using realistic deuteron wave functions~\cite{sat86};  
from descriptions of high energy SLAC electron-deuteron data 
at $x > 1.0$~\cite{yen89}; from studies of the deuteron electromagnetic 
structure functions~\cite{kobushkin}; and from calculations of fast 
backward tagged nucleons coming from deuteron breakup~\cite{chl01}.
The estimates range between 0.01 and 0.07.  We shall quote results 
assuming $f$ = 0.01 for free deuterons, and scaling $f$ appropriately
for 6q clusters within the nuclear medium. 

It will be convenient for us to represent the spin-flavor-color structure
of the deuteron 6q state in a field-theoretic form.  Letting $q^\dagger$
represent an (anticommuting) quark creation operator, the highest
projection spin state may be written

\begin{eqnarray}
\big| D, 6q, S_z=1 \big\rangle = &N& \epsilon^{\alpha\beta\gamma}
                          \epsilon^{\delta\epsilon\zeta}
              \epsilon^{ab}\epsilon^{de}\epsilon^{cf}
              \epsilon^{ij}\epsilon^{kl}  \nonumber\\
&\times&
  q^\dagger_{\alpha a i}  q^\dagger_{\beta b j}  
         q^\dagger_{\gamma c \uparrow}
  q^\dagger_{\delta d k}  q^\dagger_{\epsilon e l}  
         q^\dagger_{\zeta f \uparrow}  \big| 0 \big\rangle \ ,
\label{eq:dstate}
\end{eqnarray}

\noindent where the Greek indices represent SU(3) color, $a \ldots e$
SU(2) flavor, and $i \ldots l$ SU(2) spin, and $q^\dagger$ is a creation
operator for a quark state satisfying

\begin{equation}
\big\{ q_{\alpha a i}, q^\dagger_{\beta b j} \big\} =
    \delta_{\alpha\beta} \delta_{a b} \delta_{i j}  \ .
\end{equation}

\noindent  One may see by inspection that the
state has the quantum numbers of the deuteron. Note that the SU(2) spin
space corresponds exactly to the MIT bag model wave functions described
below (so the reader should not think that  our calculation is
nonrelativistic).  In this representation, it is  straightforward (though
tedious) to compute the normalization factor $N$; we find 
$N = 1/(48\sqrt{10})$.

\section{Matrix Element}

With the state defined, we now turn to one possible operator of interest
\begin{eqnarray}
O = [u^{T\alpha}_R C u^\beta_R ]\  [d^{T\gamma}_R C d^\delta_R] \ 
    [s^{T\rho}_R C s^\tau_R] \ \epsilon_{\alpha \gamma \rho}
                                              \epsilon_{\beta \delta \tau} \ ,
\label{eq:theop}
\end{eqnarray}
which contributes to the decay $D \rightarrow K^* K$.  Here, $C$ represents 
the charge conjugation matrix, and all the fermion fields are right-handed, 
so that $O$ is manifestly SU(3)$\times$SU(2)$\times$U(1) invariant.  
Since strange quarks are present only in the final state, the four
remaining fields annihilate quarks in the deuteron, leaving two spectators. 
We represent the quark cluster component of the deuteron as well as the 
outgoing kaons as MIT bag model states (described in more detail below), 
and define the spatial origin as the point at which our $\Delta B=2$ operator 
acts.  When quark fields are replaced by ground state bag wave functions 
multiplied by appropriate creation or annihilation operators, the 
spin-flavor-color (SFC) matrix element may be factored from the spatial one.
It may be determined by allowing
\begin{eqnarray}
\hat O = \epsilon^{\alpha \gamma \rho} \epsilon^{\beta \delta \tau}
    \epsilon^{ij} \epsilon^{kl} \epsilon^{mn}
   u_{\alpha i} u_{\beta j} d_{\gamma k} d_{\delta l} 
  \bar s^\dagger_{\rho m}  \bar s^\dagger_{\tau n}   
\end{eqnarray}
to act on the state Eq.~(\ref{eq:dstate}), where the symbols now represent
creation or annihilation operators rather than fields; one then takes the 
overlap with a similarly constructed two kaon state.  We have computed the 
SFC matrix element by hand and by symbolic mathematics 
code, and obtain
\begin{equation}
\eta_{SFC} = 
\big\langle
     K^0; K^{*+}, S_z=1
     \big| \hat O \big|    D, 6q, S_z=1 \big\rangle_{SFC}  = -4\sqrt{5}  \ .
\end{equation}

To evaluate the spatial part of the matrix element, we must
take into account that the desired external states are eigenstates of
momentum.  To relate these to bag states, we suppose that the
momentum eigenstates form a complete set, so that~\cite{cc81}

\begin{eqnarray}
|D(\vec R) \rangle_B = \int {d^3 P \over (2\pi)^3 2 E }
              \phi(\vec P) e^{i\vec P \cdot \vec R} 
              |D(\vec P) \rangle  \ ,
\end{eqnarray}

\noindent and similarly for the other states.  The state with the
subscript ``B'' is a bag state centered at $\vec R$, spin variables are
tacit, and the momentum eigenstate is normalized by

\begin{eqnarray}
\langle D(\vec P') | D(\vec P) \rangle = 
       2E (2\pi)^3 \delta^3 (\vec P - \vec P') \ .
\end{eqnarray}

\noindent By inversion, 

\begin{eqnarray}
\big| D(\vec P) \big\rangle = {2E \over \phi(\vec P) } 
         \int d^3 R  \ e^{-i\vec P \cdot \vec R} 
         \big| D(\vec R) \big\rangle_B  \ .
\end{eqnarray}

\noindent The normalization condition determines 
$\phi(P) \equiv \phi(\vec P)$,

\begin{eqnarray}
\phi^2 (P) = 2E \int d^3 r e^{-i\vec P \cdot \vec r} I_n(r) 
      = 2E \widetilde I_n(P)\ ,
\end{eqnarray}

\noindent where $I_n$ is the wave function overlap of two $n$-quark
states centered at different points,

\begin{eqnarray}
I_n(r) = {\lower.9ex\hbox{ $\scriptstyle B$}\big\langle} 
            D(-{1\over 2} \vec r) \big| 
                D({1\over 2} \vec r) \big\rangle_B  \ .
\end{eqnarray}

\noindent For applications where the wave function is written as the
product of $n$ independent quark wave functions, one has

\begin{equation}
I_n(r) = \big( I_1(r) \big)^n \ ,
\end{equation}

\noindent with
 
\begin{eqnarray}
I_1(r) \equiv {\lower.9ex\hbox{ $\scriptstyle B$}\big\langle} 
            q(-{1\over 2} \vec r) \big| 
                q({1\over 2} \vec r) \big\rangle_B  \ .
\end{eqnarray}
In the bag model, this may be written explicitly as 
\begin{equation}
I_1(r)= (4 \pi) \int_0^{R_B-r/2} dz \int_0^{\sqrt{R_B^2-(z+r/2)^2}}
\kern -2em \rho d\rho (u_+ u_- + l_+ l_- \hat R_+ \cdot \hat R_- )
\end{equation}
where,
\begin{equation}
\begin{array}{ccc}
R_\pm = [(r/2 \mp z)^2+\rho^2]^{1/2} & \mbox{ , } & 
\hat R_+ \cdot \hat R_- = (R_+^2+R_-^2-r^2)/(2R_+R_-)
\end{array}
\end{equation}
and where the factors $u_i = u(R_i)$ and $l_i = l(R_i)$ are the upper and
lower components of bag wave functions 
\begin{eqnarray}
\psi(\vec r) =    \left( 
  \begin{array}{c}
    u(r) \chi  \\[1.0ex]
   i l(r) \vec\sigma \cdot \hat r \chi  
  \end{array}
                   \right)  =
  {N\over \sqrt{4\pi}}     \left( 
  \begin{array}{c}
     j_0(\omega r) \chi  \\[1.0ex]
   i j_1(\omega r) \vec\sigma \cdot \hat r \chi  
  \end{array}
                   \right)  \,\,\, .
\end{eqnarray}
Here $\chi$ is a Pauli spinor, $\omega = x/R_B = 2.043/R_B$,  and
$N^2 R_B^3 = x/[2(x-1) j_0^2(x)]\approx 5.15$.

To calculate the matrix element for the decay 
$D(\vec P = 0) \rightarrow K^*(\vec p) K(-\vec p)$, we apply the
foregoing formalism to the quark cluster part of the deuteron as well as
to the kaons.  For simplicity, we assume the same bag radius $R_B\approx1$~fm 
for all states.  We obtain

\begin{eqnarray}   \label{eq:mtilde}
{\widetilde M} &=& \big\langle K(-\vec p); K^*(\vec p), S_z=1
                 \big|  O \big|
           D(\vec P = 0), 6q, S_z=1 \big\rangle  \nonumber \\
&=& \left[ 2m_D 2E_1 2E_2 \over \widetilde I_6(0) \widetilde I_2^{\ 2}(p)
                                             \right]^{1/2}
                                                \times \widetilde N
\end{eqnarray}

\noindent and

\begin{eqnarray}   \label{ME}
\widetilde N &=& 
 \int d^3 R \, d^3 R_1 \, d^3 R_2 \ e^{ip(Z_1 - Z_2)}  \nonumber \\
&\times&
  {\lower.9ex\hbox{ $\scriptstyle B$}\big\langle} 
     K(\vec R_1); K^*(\vec R_2), S_z=1
     \big|  O \big|    D(\vec R), 6q, S_z=1 \big\rangle_B  \ ,
\end{eqnarray}

\noindent where $E_1$ and $E_2$ are the energies of the $K$ and $K^*$,
respectively, and the $\hat z$-direction is taken parallel to $\vec p$.
This may be re-expressed as
\begin{eqnarray}
\widetilde N &=&  {1\over 8}
                        \eta_{SFC} \int d^3 R \, d^3 R_1 \, d^3 R_2 \ 
e^{ip(Z_1 - Z_2)}  
                \nonumber  \\[1ex]
&\times& I_1(r_{10}) I_1(r_{20}) (u^2+l^2)^2 
               (u_1 u_2 + l_1 l_2 \hat R_1 \cdot \hat R_2 ) \ ,
\end{eqnarray}
where $u=u(R)$, $l=l(R)$, and $r_{i0} = |\vec R_i - \vec R|$.  The factors 
of $I_1$ come from the overlap of spectator quark wave functions.  All the 
integrals in Eq.~(\ref{eq:mtilde}) may be evaluated numerically. For
$p R_B=3.12$, we find 
$\widetilde N = \eta_{SFC} (N^2 R_B^3/8\pi)^3 \times 0.31$,
$\widetilde I_2(p) = R_B^3 \times 0.41$, and
$\widetilde I_6(0) = R_B^3 \times 0.13$.

%%%%%%%%%%%%%%%%%%%%%%%%%%%%%%%%%%%%%%%%%%%%

 \section{Results}

%%%%%%%%%%%%%%%%%%%%%%%%%%%%%%%%%%%%%%%%%%%%
Using the results from the previous section, it is straightforward to derive 
the decay width.  In keeping with the proton decay literature, we will
instead express our result in terms of a `partial lifetime' ({\em i.e.}
the lifetime if the branching fraction to the given mode were 100\%).
We find
\begin{equation}
\tau (D \rightarrow K^* K) = 2.18 \times 10^7 \mbox{ yrs } \cdot \left(\frac{M}{1\mbox{ TeV}} \right)^{10} f^{-1}  \,\,\, 
\end{equation}
where $M^5$ is the dimensionful factor that suppresses the dimension-nine 
operator of interest. (While we identify $M$ with the Planck scale in the
present discussion, it is worth mentioning that our calculational framework
is applicable to any model, {\em e.g.} R-parity violating supersymmetry,
in which such effective operators are induced.)  For a free deuteron (for 
example, in $D_2 O$), we set $f=0.01$ and find $\tau \sim 10^{34}$~years 
for $M\sim 293$~TeV. For a deuteron within the O$^{16}$ nucleus, the quark 
cluster probability will scale as the nuclear density; for a reasonable 
estimate, $f=0.2$, one finds $\tau \sim 10^{34}$~years for $M\sim 395$~TeV.  
Varying $M$ by a factor of $2$ changes the lifetime by $2^{10}\sim 10^3$, 
which allows the result to vary from well below to significantly above the 
lifetimes usually associated with the maximum proton decay reach of 
Super-Kamiokande.

A more realistic estimate of the decay rate would also include flavor
suppression factors that provide a small dimensionless number multiplying
our operator.  Without a model of flavor, however, we cannot exclude the
extreme possibility in which Eq.~(\ref{eq:theop}) has no further suppression
and all other $\Delta B=2$ operators are simply absent.  On the other hand,
we may consider how our results change if we impose a simple flavor ansatz:
we could assume that fields of the first (second) generation are suppressed
by a factor of $\lambda^3$ ($\lambda^2$), where $\lambda \sim 0.2$ is of
comparable size to the Cabibbo angle. (This could arise if a $\Delta B=2$ 
operator involving only third generation fields is allowed, and all others are
generated from it via CKM-like rotations on the fields~\cite{fn1}.)  In 
this case, our previous estimate for $\tau$ is extended by $\lambda^{-32} \sim
10^{22}$; for decays in O$^{16}$ we now find $\tau \sim 10^{34}$~years
for $M=2.3$~TeV.  With this ansatz, we can now also say something about
the strangeness conserving operators that contribute to $n$-$\overline{n}$
oscillation. Estimates of the relevant matrix elements already
exist~\cite{shrock}, from which we find
\begin{equation}
\tau \sim 10^{-12} \mbox{ yrs } \cdot 
\left(\frac{M}{1\mbox{ TeV}}\right)^5 \lambda^{-18}  = 7.6\mbox{ yrs.} \cdot 
\left(\frac{M}{1\mbox{ TeV}}\right)^5
\end{equation}
which exceeds the best experimental bound, $3.8$~yrs., 90\% CL \cite{RPP}
for any value of $M$ that could be plausibly identified with the Planck
scale.

Table~1 shows other possible nucleon-nucleon decay modes, and the 
multiplicative correction that must be applied to our $D\rightarrow K K^*$
result to obtain an estimate for the lifetime.  This factor takes into
account the differing phase space, spatial wave function overlaps
and flavor suppression following from our previous ansatz.  We simply
assume the SFC matrix elements are of comparable size. While we don't compute 
these matrix elements explicitly (given the far larger uncertainty from
the proliferation of operator coefficients) the contribution of any 
given operator of interest may be evaluated explicitly using the approach 
we have presented. Observation of any of these modes in the absence of 
conventionally expected nucleon decays would be a remarkable and unexpected 
sign of exotic physics not too far beyond directly accessible energies.

{\samepage
\begin{center}
{\bf Acknowledgments}
\end{center}
We thank the National Science Foundation for support under Grant No.\ 
PHY-9900657. In addition, C.D.C. thanks the Jeffress Memorial Trust 
for support under Grant No.~J-532.  
}
%\appendix

\begin{table}
\begin{center}
\begin{tabular}{cccccc}
$\Delta S =0$ &$(\times \lambda^{-4})$ & $\Delta S =1$ & 
$(\times \lambda^{-2})$ & $\Delta S=2$ & $(\times \lambda^0)$  \\ \hline\hline
$\pi\pi$    & 2.54 &  $K\pi$     & 1.91 &  $KK^*$    & 1.00 \\
$\rho\pi$   & 1.40 &  $K^*\pi$   & 1.25 &  $KK$      & 1.49 \\
$\rho\rho$  & 0.90 &  $K\rho$    & 1.11 &  $K^*K^*$  & 1.10 \\
$\pi\eta$   & 1.85 &  $K^*\rho$  & 0.90 &            & \\
$\rho\eta$  & 1.06 &  $K\eta$    & 1.40 &            & \\
$\pi\eta'$  & 1.18 &  $K^*\eta$  & 0.96 &            & \\
$\rho\eta'$ & 0.96 &  $K\eta'$   & 0.96 &            & \\ 
$\eta\eta'$ & 0.94 &  $K^*\eta'$ & 1.87 &            & \\
$\eta\eta$  & 1.32 &             &      &            & \\
\hline
\end{tabular}
\caption{Lifetime correction factors for typical $NN$ decay modes.
The numbers do not include the flavor parameter $\lambda \approx 0.2$,
which should be multiplied in as indicated.}
\end{center}
\end{table}


\begin{thebibliography}{99}
\bibitem{fn1}
We assume that the $\Delta B=0$ flavor changing operators
are either absent or highly suppressed, though we do not
attempt to address this model-dependent issue directly.  For 
related work, see Ref.~\cite{fcncs}.
\end{thebibliography}
\end{document}